\documentclass[review, sort&compress]{elsarticle}
\usepackage{epsfig,amsmath,graphicx,amssymb,overpic}
\usepackage{bm}
\usepackage{mathrsfs}
\usepackage{graphicx}
\usepackage{epsfig}
\usepackage{dcolumn}
\usepackage{bm}
\usepackage{amsmath,amssymb,amsthm}
\usepackage[colorlinks=true,linkcolor=blue]{hyperref}
\usepackage{subfigure}
\usepackage{booktabs}
\usepackage[mathscr]{euscript}
\usepackage{ulem}
\usepackage{setspace}
\usepackage{color}
\usepackage{geometry}  
\newcommand{\bea}{\begin{eqnarray}}
\newcommand{\eea}{\end{eqnarray}}
\newcommand{\beq}{\begin{equation}}
\newcommand{\eeq}{\end{equation}}

\def\/{\over}

\usepackage{CJK}

\begin{document}
	
\title{
From infinite symmetries to multi-soliton manifolds: dispersion-driven reduction for the KP and mKP equations}

\author[rvt1]{Xiazhi Hao}
\author[rvt1]{Mingxia Ji}%
\author[rvt2,rvt3]{S. Y. Lou\corref{cor1}}
\cortext[cor1]{Corresponding author. Email: lousenyue@nbu.edu.cn(S. Y. Lou) }

\address[rvt1]{School of Mathematical Sciences, Zhejiang University of Technology, Hangzhou, 310014, China}
\address[rvt2]{School of Physical Science and Technology, Ningbo University, Ningbo, 315211, China}
\address[rvt3]{Institute of Fundamental Physics and Quantum Technology, Ningbo University, Ningbo, 315211, China}

\begin{abstract}	
The infinite-dimensional symmetry algebras of integrable systems in $(2+1)$ dimensions exhibit a remarkable reduction property when restricted to finite-dimensional multi-soliton manifolds. In this paper, we systematically investigate this phenomenon for the Kadomtsev-Petviashvili   and modified Kadomtsev-Petviashvili equations. For the Kadomtsev-Petviashvili equation, we analyze both the $K$-symmetry  and the $\tau$-symmetry hierarchies. It is shown that for any $n$-soliton solution, each higher-order $K$-symmetry generator reduces to a linear combination of the $n$ independent translational symmetries $\sigma_{\xi_{m0}}$ associated with the individual soliton phase shifts, where the coefficients are precisely the dispersion relations evaluated at the soliton parameters. Closed-form expressions for odd- and even-order $K$-symmetries are derived, and determinant representations are provided to characterize the reduction systematically. 
For the \(\tau\)-symmetries, each \(\tau_n\) becomes a linear combination of the three fundamental soliton symmetries \(\sigma_{k_m}\), \(\sigma_{l_m}\), and \(\sigma_{\xi_{m0}}\), with coefficients that are rational functions of soliton parameters \(k_m\) and \(l_m\). 
These results are further extended to the modified Kadomtsev-Petviashvili equation, where analogous reduction patterns are established.
This work establishes a direct and intrinsic link between the algebraic structure of infinite symmetries and the spectral properties of soliton solutions, offering a unified physical interpretation: the infinite set of higher-order symmetries generates linear superpositions of elementary soliton translations, weighted by powers of the wave numbers and dispersion coefficients. 
\end{abstract}

\maketitle
\large
\doublespacing
\section{Introduction}

The theory of nonlinear integrable systems is remarkable for its ability to unify seemingly disparate mathematical structures: infinite-dimensional Lie algebras, algebraic curves, Riemann-Hilbert problems, and tau functions, all within a harmonious framework \cite{jimbo1983solitons}. Since the 1960s, from the Korteweg-de Vries (KdV) equation to the Kadomtsev-Petviashvili (KP) equation, symmetry analysis has served as a fundamental tool for revealing the intrinsic geometry of these systems \cite{rbib23,fuchssteiner1981kdv}. 
For an integrable partial differential equation
\[
u_t = K(u), \quad (x,t) \in \mathbb{R}^{d+1},
\]
an infinitesimal symmetry generator \(\sigma\) is characterized by the linearized equation 
\[
\sigma_t = K'(u)[\sigma],
\]
where \(K'(u)[\sigma] \equiv \left.\frac{d}{d\varepsilon} K(u+\varepsilon \sigma)\right|_{\varepsilon=0}\) denotes the Fr\'echet derivative of the nonlinear operator \(K\) in the direction \(\sigma\). For \((1+1)\)-dimensional systems, the existence of a recursion operator \(\Phi\) allows the entire infinite hierarchy of commuting symmetries to be compactly represented as \(K_n = \Phi^n u_x\), thereby generating the celebrated integrable hierarchy \cite{FokasFuchssteiner1981, wbib28, pbib7, bbib5, bbib6}.

However, upon entering the realm of (2+1) dimensions, the situation changes fundamentally. A universal recursion operator generally ceases to exist, rendering the construction of symmetries a substantially more challenging problem. Nevertheless, several powerful approaches have been developed to address this difficulty. 
Date, Jimbo, Kashiwara, and Miwa \cite{Date1981KP} systematically constructed the KP hierarchies of orthogonal and symplectic type using the fermionic operator formalism based on the Clifford algebra, where the symmetries are generated by infinite-dimensional Lie groups \(\mathrm{O}(\infty)\) and \(\mathrm{Sp}(\infty)\), respectively.
In parallel, Chen, Lee, and Lin \cite{hbib9} discovered a new hierarchy of symmetries for the KP equation that depend explicitly on space-time variables. These symmetries, together with the classical ones, form an infinite-dimensional Lie algebra, with the commutation relations derivable via a geometric analysis of their action on soliton solutions.
Fuchssteiner \cite{bbib7} formalized these structures through the notion of mastersymmetries that map the commutant of a given flow into itself, thereby providing a unified algebraic framework for constructing both time-independent and time-dependent symmetries.
Subsequently, Cheng \cite{ybib23} developed a Lax-operator algebraic approach for (2+1)-dimensional integrable systems including modified KP (mKP) and Caudrey-Dodd-Gibbon-Katera-Sawada, showing that the hierarchies of isospectral and non-isospectral flows, along with their symmetries and mastersymmetries, follow directly from the Lie-algebraic structure of the Lax operators. One of us, Lou \cite{lou1993,sbib20} extended the mastersymmetry method by incorporating arbitrary functions of time into the symmetries of the KP equation, and derived an explicit constructive formula for the entire hierarchy of generalized symmetries. Through these complementary approaches, infinitely many commuting symmetries  have been constructed for the KP equation and its related hierarchies, generating infinite-dimensional Lie algebras.

Despite the completeness of the algebraic construction, a clear and unified physical or geometric interpretation of the action of these infinitely many abstract symmetries on the solution manifold has been conspicuously absent. The key observation is elementary yet profound: while the equation itself possesses an infinite-dimensional symmetry algebra, its most physically significant exact solutions, the \(n\)-soliton solutions, reside on a finite-dimensional submanifold \(\mathcal{M}_n\) parametrized by a finite set of \(3n\) parameters, namely,
the wave numbers \(k_m\), the direction parameters \(l_m\), and the phase shifts \(\xi_{m0}\).

This stark dimensional disparity necessitates a deep reduction phenomenon: upon restriction to the finite-dimensional soliton submanifold \(\mathcal{M}_n\), the infinite-dimensional symmetry algebra must degenerate, its rank collapsing to a finite number \cite{lou1, mbib26, mbib28, 25pla, 26zamp}. 
This reduction phenomenon is not entirely new: for (1+1)-dimensional integrable systems possessing a hereditary recursion operator, Fuchssteiner and Oevel \cite{bbib16} showed that multi-soliton manifolds are reductions with respect to suitable linear combinations of symmetry generators, whose tangent space is spanned by parameter derivatives that are eigenvectors of the recursion operator. The novelty of the present work is that the same mechanism persists in (2+1) dimensions, where no universal recursion operator exists, with the dispersion relations playing the role of the spectral multipliers.
More precisely, there exists a natural restriction map from the infinite symmetry algebra to the tangent space, which is spanned by the independent translational generators, corresponding to the individual phase shifts of each soliton. 
The central problem addressed in this paper is therefore the following: given an infinite-dimensional symmetry algebra associated with an integrable hierarchy, what is its precise action on the finite-dimensional soliton submanifold? To this end, we shall develop a systematic framework for computing the restriction of symmetries to \(n\)-soliton solutions.

The investigation proceeds along two main lines. First, for the KP equation, we derive explicit reduction formulas for  \(K\)-symmetries and \(\tau\)-symmetries on the \(n\)-soliton manifold, and identify the resulting algebraic structure of the restricted symmetry algebra. Second, we extend the analysis to the mKP equation, in order to test the universality of the reduction mechanism across different Lax representations and dispersion relations.

\section{Infinite symmetries of the KP equation and their reduction on multi-soliton manifolds}\label{kpsyms}

We consider the KP equation in the form
\begin{eqnarray}\label{kp}
u_t=u_{xxx}+6uu_x-3\partial_{x}^{-1}u_{yy},
\end{eqnarray}
which models weakly dispersive and weakly nonlinear two-dimensional wave propagation \cite{Kadomtsev1970, sbib15}. Here \(\partial_{x}^{-1}\) denotes the formal inverse of \(\partial_{x}\), normalized by \(\partial_{x}\partial_{x}^{-1}=\partial_{x}^{-1}\partial_{x}=1\).
A vector field \(\sigma(x,t,u)\) is called an infinitesimal symmetry generator of equation \eqref{kp} when the one-parameter transformation \(u \mapsto u + \epsilon\sigma + O(\epsilon^2)\) preserves its solution set. The symmetry generator obeys the linearized evolution equation
\begin{eqnarray}\label{le}
\sigma_t=K'[\sigma]=\sigma_{xxx}+6u\sigma_x+6u_x\sigma-3\partial_x^{-1}\sigma_{yy}.
\end{eqnarray}

\subsection{Infinite $K$- and $\tau$-symmetries}
The linearized equation \eqref{le} for infinitesimal symmetries of the KP equation admits an infinite family of solutions. These symmetry generators fall into two separate infinite hierarchies, conventionally known as the \(K\)-symmetries and the \(\tau\)-symmetries.
The \(K\)-symmetries are time-independent and commute among themselves, forming the classical symmetry algebra of the KP hierarchy. The \(\tau\)-symmetries, by contrast, depend explicitly on the space-time variables \((x,y,t)\) and generate a non-abelian algebra that extends the classical one.

An explicit construction of the commuting \(K\)-symmetries was given in \cite{OevelFuchssteiner1982,sbib30,sbib20,22mmas}
\[\sigma_n=-\frac{1}{2(n-1)!3^n}L^ny^{n-1},~n=1,2,...,\infty,\]
with 
\[L\equiv \partial_x^3+6\partial_xu-3\partial_x^{-1}\partial_y^2-\partial_t.\]
The first few members of this hierarchy are
	\begin{eqnarray}\begin{split}\label{ks}
		&\text{KP1}: u_{t_1}=K_1=-u_x,\\
		&\text{KP2}: u_{t_2}=K_2=2u_y,\\
		&\text{KP3}: u_{t_3}=K_3=u_{xxx}+6uu_x-3\partial^{-1}_xu_{yy},\\
		&\text{KP4}: u_{t_4}=K_4=-4(u_{xxy}+4uu_y+2u_x\partial^{-1}_xu_{y}-\partial^{-2}_{x}u_{yyy}),\\
		&\text{KP5}: u_{t_5}=K_5=-u_{5x}-5\partial^{-3}_xu_{4y}+10(u_{xyy}-uu_{xxx}-2u_xu_{xx}-3u^2u_x)\\&\qquad\qquad\qquad\quad+10(u_x\partial^{-2}_xu_{yy}+\partial^{-1}_xuu_{yy}+\partial^{-1}_xu_y^2+2u\partial^{-1}_xu_{yy}+2u_y\partial^{-1}_xu_y),\\
		&\text{KP6}: u_{t_6}=K_6=-20u_{yyy}+6(\partial^{-4}_xu_{5y}+u_{4xy}+12u_xu_{xy}+16u^2u_y+8u_{xx}u_y\\&\qquad\qquad\qquad\quad+8uu_{xxy})+12(u_{xxx}\partial^{-1}_xu_y-\partial^{-1}_x(u_y\partial^{-1}_xu_{yy})-\partial^{-1}_x(u\partial^{-1}_xu_{yyy})\\&\qquad\qquad\qquad\quad-\partial^{-2}_xuu_{yyy}-u_x\partial^{-3}_xu_{yyy}-2u_y\partial^{-2}_xu_{yy}-2u\partial^{-2}_xu_{yyy}+2u_x\partial^{-1}_xuu_y\\&\qquad\qquad\qquad\quad-3(\partial^{-1}_xu_y)(\partial^{-1}_xu_{yy})-3\partial^{-2}_xu_yu_{yy}+6uu_x\partial^{-1}_xu_y).
		\end{split}
	\end{eqnarray}
When $u_y=0$, the KP hierarchy is reduced to the KdV hierarchy.
Like those in other integrable systems, these symmetries do not depend on the space and the time variables explicitly. They are related to the classical conserved densities which can be derived from the Lax pair of operators for equation \eqref{kp}. 
The lowest-order symmetries \(K_{1},K_{2},K_{3}\) correspond respectively to translations in \(x\), \(y\) and \(t\), with \(K_{3}\) being precisely the right-hand side of equation \eqref{kp}. The entire hierarchy satisfies \([K_{m},K_{n}]=0\) for all \(m,n\geq 1\).

The \(\tau\)-symmetries,  identified in \cite{hbib9,hbib1}, depend explicitly on \(x,y,t\). Their construction proceeds via the recursion
\begin{align}\label{tn}
\tau_{n+1}=\frac{1}{n-5} [\tau_4, \tau_n], ~ n>5
\end{align} 
with \(\tau_{4}\) given by

\begin{eqnarray*} 
	&&\tau_4=-12t(u_{xxy}+4uu_y+2u_x\partial^{-1}_xu_{y}-\partial^{-2}_{x}u_{yyy})+y(u_{xxx}+6uu_x-3\partial_{x}^{-1}u_{yy})-2xu_y-4\partial_x^{-1}u_y\\&&~~~=3tK_4+yK_3-xK_2-4\partial_x^{-1}u_y.
\end{eqnarray*}
The first three $\tau$-symmetries are
\begin{eqnarray*}
&&\tau_1=-3tu_x-\frac12=3tK_1-\frac12,\\
&&\tau_2=6tu_y-yu_x=3tK_2+yK_1,\\
&&\tau_3=3t(u_{xxx}+6uu_x-3\partial_{x}^{-1}u_{yy})+2yu_y+xu_x+2u=3tK_3+yK_2-xK_1+2u.
\end{eqnarray*}
They correspond to the Galilean invariance, a special rotational invariance, and the scaling invariance of the KP equation \eqref{kp}, respectively. 
We present three more $\tau$-symmetries 
\begin{align*}
\tau_5&=3tK_5+yK_4-xK_3+6\partial_x^{-2}u_{yy}-8u^2-4u_{xx}-2u_x\partial_x^{-1}u,\\
\tau_6&=3tK_6+yK_5-xK_4+4(4u_{xy}+6u\partial_x^{-1}u_y+u_x\partial_x^{-2}u_y+u_y\partial_x^{-1}u+4\partial_x^{-1}(uu_y)-2\partial_x^{-3}u_{yyy}),\\
\tau_7
&= 3tK_7 + yK_6 - xK_5 + 2\Big(
16u^3 + 18u_x^2 + 24uu_{xx} + 3u_{xxxx} - 20u_{yy}
\Big) \\
&+ 2\Big(
6uu_x\partial_x^{-1}u
+ 3u_x\partial_x^{-1}u^2
- 3u_x\partial_x^{-3}u_{yy}
- 16u\partial_x^{-2}u_{yy}
- 14\big(\partial_x^{-1}u_y\big)^2 \\
&\quad
- 4u_y\partial_x^{-2}u_y
- 13\partial_x^{-1}u_{yy}\partial_x^{-1}u
+ u_{xxx}\partial_x^{-1}u
- 10\partial_x^{-2}(uu_{yy})
+ 5\partial_x^{-4}u_{yyyy} \\
&\quad
- 10\partial_x^{-2}u_y^2
+ 10\partial_x^{-1}\big(u_{yy}\partial_x^{-1}u \big)
\Big).
\end{align*}
The classical \(K\)-symmetries \eqref{ks} can also be recovered from the \(\tau_4\) via \cite{sbib56}
\begin{eqnarray*}\label{kn}
K_{n+1}=\frac{1}{n} [\tau_4, K_n], \quad n \geq 1,\quad  K_1=-u_x.
\end{eqnarray*} 
To analyze the reduction properties of these symmetries on soliton background, we first present the $n$-soliton solutions of the KP equation \eqref{kp}.

\subsection{Multi-soliton solutions}

The KP equation \eqref{kp} admits the standard \(n\)-soliton solutions expressed in Hirota form \cite{BiondiniChakravarty2006, ChakravartyKodama2008}
\begin{eqnarray}\label{ns}
u_n=2(\ln f_n)_{xx},
\end{eqnarray}
where \(f_n\) is defined as a sum over all binary-valued index sets \(\mu=\{\mu_i|i=1,2,\dots,n\},\mu_i\in\{0,1\}\)
\begin{eqnarray*}
f_n= \sum_{\mu}\exp\left(\sum_{j=1}^n\mu_j\xi_j+\sum_{1\leq j<p\leq n}\mu_j\mu_p\theta_{jp}\right),
\end{eqnarray*}
with
$$\xi_j=k_jx+l_jy+\displaystyle\frac{k_j^4-3l_j^2}{k_j}t+\xi_{j0},~\exp(\theta_{jp})=\displaystyle\frac{k_j^2k_p^2(k_j-k_p)^2+(k_jl_p-k_pl_j)^2}{k_j^2k_p^2(k_j+k_p)^2+(k_jl_p-k_pl_j)^2}.$$
Explicitly, the
	one-, two- and three-soliton solutions of equation \eqref{kp} take the form 
	\begin{eqnarray}\label{kp1}
	u_{1}&=&2(\ln f_1)_{xx},~ f_1=1+\exp(\xi_1),\\\label{kp2}
	u_{2}&=&2(\ln f_2)_{xx},~ f_2=1+\exp(\xi_1)+\exp(\xi_2)+A_{12}\exp(\xi_1+\xi_2),\\\label{kp3}
	u_{3}&=&2(\ln f_3)_{xx},~ f_3=1+\exp(\xi_1)+\exp(\xi_2)+\exp(\xi_3)+A_{12}\exp(\xi_1+\xi_2)\\\nonumber&&\qquad\qquad\qquad+A_{13}\exp(\xi_1+\xi_3)+A_{23}\exp(\xi_2+\xi_3)\\\nonumber&&\qquad\qquad\qquad+A_{12}A_{13}A_{23}\exp(\xi_1+\xi_2+\xi_3),
	\end{eqnarray}
	where
	$$~\xi_j=k_jx+l_jy+\displaystyle\frac{k_j^4-3l_j^2}{k_j}t+\xi_{j0},\qquad j=1,2,3,$$ and
	$$A_{ij}=\displaystyle\frac{k_i^2k_j^2(k_i-k_j)^2+(k_il_j-k_jl_i)^2}{k_i^2k_j^2(k_i+k_j)^2+(k_il_j-k_jl_i)^2},\qquad 1\leq i<j\leq 3.$$	
The $n$-soliton manifold $M_n$ is a finite‑dimensional submanifold embedded in the infinite‑dimensional solution space of the KP equation. It is parametrized by $3n$ independent real parameters $\{k_m,l_m,\xi_{m0}\}_{m=1,2,\dots,n}$. Accordingly, its tangent space $T_{u_n}M_n$ at an $n$-soliton solution $u_n$ is spanned by three families of vector fields corresponding to variations of each soliton parameter
\begin{equation*}
\sigma_{k_m}=\frac{\partial u_n}{\partial k_m},\quad
\sigma_{l_m}=\frac{\partial u_n}{\partial l_m},\quad
\sigma_{\xi_{m0}}=\frac{\partial u_n}{\partial \xi_{m0}},\qquad m=1,2,\dots,n.
\end{equation*}
As we shall demonstrate in the following subsection, the entire infinite \(K\)-symmetry hierarchy collapses on this \(n\)-dimensional subspace, with each \(K_{i}\) reducing to a linear combination of the \(\sigma_{\xi_{m0}}\) terms, the coefficients of which are precisely the dispersion relations evaluated at the soliton parameters.

\subsection{Restrictions of higher‑order symmetries on multi-soliton solutions}\label{kpkt}

In this subsection, we systematically investigate symmetry transformations associated with the KP equation by computing their explicit actions on $n$-soliton solutions. For the $n$-soliton solution \eqref{ns}, the independent symmetries $\sigma_{\xi_{m0}}$ ($m = 1, 2, \dots, n$) are known to generate translations of the individual soliton phase shifts. We show that each higher-order symmetry $K_i$ reduces to a linear combination of these fundamental symmetries, with coefficients given explicitly in terms of the soliton parameters $k_m$ and $l_m$. Specifically, the restrictions of the first several $K_i$ take the following forms
\begin{eqnarray*}
	&&K_1|_{u=u_n}=-\sum_{m=1}^nk_mu_{n,\xi_{m0}}=-\sum_{m=1}^nk_m\sigma_{\xi_{m0}},\\
	&&K_2|_{u=u_n}=2\sum_{m=1}^nl_mu_{n,\xi_{m0}}=2\sum_{m=1}^nl_m\sigma_{\xi_{m0}},\\
	&& K_3|_{u=u_n}=\sum_{m=1}^n\frac{k_m^4-3l_m^2}{k_m}u_{n,\xi_{m0}}=\sum_{m=1}^n\frac{k_m^4-3l_m^2}{k_m}\sigma_{\xi_{m0}},\\
	&&K_4|_{u=u_n}=-2\sum_{m=1}^n\frac{2k_m^4-2l_m^2}{k_m^2}l_mu_{n,\xi_{m0}}=-2\sum_{m=1}^n\frac{2k_m^4-2l_m^2}{k_m^2}l_m\sigma_{\xi_{m0}},\\
	&&K_5|_{u=u_n}=-\sum_{m=1}^n\frac{k_m^8-10k_m^4l_m^2+5l_m^4}{k_m^3}u_{n,\xi_{m0}}=-\sum_{m=1}^n\frac{k_m^8-10k_m^4l_m^2+5l_m^4}{k_m^3}\sigma_{\xi_{m0}},\\
	&&K_6|_{u=u_n}=2\sum_{m=1}^n\frac{3k_m^8-10k_m^4l_m^2+3l_m^4}{k_m^4}l_mu_{n,\xi_{m0}}=2\sum_{m=1}^n\frac{3k_m^8-10k_m^4l_m^2+3l_m^4}{k_m^4}l_m\sigma_{\xi_{m0}},\\
	&&K_7|_{u=u_n}=\sum_{m=1}^n\frac{k_m^{12}-21k_m^8l_m^2+35{k_m^4}l_m^4-7l_m^6}{k_m^5}u_{n,\xi_{m0}}=\sum_{m=1}^n\frac{k_m^{12}-21k_m^8l_m^2+35{k_m^4}l_m^4-7l_m^6}{k_m^5}\sigma_{\xi_{m0}},\\
	&&K_8|_{u=u_n}=-2\sum_{m=1}^n\frac{4k_m^{12}-28k_m^8l_m^2+28k_m^4l_m^4-4l_m^6}{k_m^6}l_mu_{n,\xi_{m0}}\\&&\qquad\quad~=-2\sum_{m=1}^n\frac{4k_m^{12}-28k_m^8l_m^2+28k_m^4l_m^4-4l_m^6}{k_m^6}l_m\sigma_{\xi_{m0}},\\
	&&\qquad\qquad \vdots
\end{eqnarray*}
The above system explicitly demonstrates that for an $n$-soliton solution $u = u_n$, each symmetry flow $K_i$ reduces to a linear combination of the independent symmetries $\sigma_{\xi_{m0}}$. In detail, one has
\[
K_i\big|_{u=u_n} = \sum_{m=1}^{n} \Omega_i(k_m, l_m) \, \sigma_{\xi_{m0}},
\]
where the coefficients $\Omega_i(k_m, l_m)$ are precisely the dispersion relations of the soliton solution for the flow $u_t = K_i$.

For the KP hierarchy, the dispersion relation for the $i$-th flow is determined by the linearized dispersion law $\omega_i = \Omega_i(k_m, l_m)$, which depends on the parameters $k_m$ and $l_m$ of the $m$-th soliton. Consequently, each symmetry $K_i$ acts on the $n$-soliton solution by weighting the individual soliton symmetries $\sigma_{\xi_{m0}}$ with the corresponding dispersion coefficient $\Omega_i(k_m, l_m)$. This reveals a direct and intrinsic relationship between the algebraic structure of the symmetries and the spectral properties of the soliton solutions, namely that the coefficients appearing in the restriction of $K_i$ to the soliton subspace are exactly the dispersion functions of the linearized hierarchy evaluated at the soliton parameters.
Thus, the reduction of higher-order symmetries to the finite-dimensional manifold of $n$-soliton solutions is governed by the dispersion relations, which act as spectral multipliers linking the abstract symmetry generators to the physically observable translational modes of individual solitons.

Based on the above reduction patterns for the first several $K_i$, we observe that for the $n$-soliton solution $u = u_n$, the restrictions of the higher-order KP symmetries $K_i$ to the finite-dimensional soliton manifold can be uniformly expressed in terms of the independent symmetries $\sigma_{\xi_{m0}}$. After a systematic analysis, we obtain the following compact closed-form expressions for the odd-order and even-order symmetries, which are valid for all positive integers $i \ge 1$ 
\begin{eqnarray}\nonumber
&&K_{2i-1}|_{u=u_n}=\sum_{m=1}^n\frac{{\rm i}}{2k_m^{2i-1}}\left[({\rm i}k_m^2+l_m)\left(-\frac{(k_m^4+l_m^2)^2}{(k_m^2+{\rm i}l_m)^2}\right)^{i-1}\right.\\\nonumber
&&\qquad\qquad~~\left.+({\rm i} k_m^2-l_m)\left(-\frac{(k_m^4+l_m^2)^2}{(k_m^2-{\rm i}l_m)^2}\right)^{i-1}\right]u_{n,\xi_{m0}}\\\label{ko}
&&\qquad\qquad~~=\sum_{m=1}^n\frac{{\rm i}}{2k_m^{2i-1}}\left[({\rm i}k_m^2+l_m)\left(-\frac{(k_m^4+l_m^2)^2}{(k_m^2+{\rm i}l_m)^2}\right)^{i-1}\right.\\\nonumber
&&\qquad\qquad~~\left.+({\rm i} k_m^2-l_m)\left(-\frac{(k_m^4+l_m^2)^2}{(k_m^2-{\rm i}l_m)^2}\right)^{i-1}\right]\sigma_{\xi_{m0}},\ {\rm i}^2=-1, \\\nonumber
&&K_{2i}|_{u=u_n}=\sum_{m=1}^n (-1)^{i+1}\frac{{\rm i}[(k_m^2+{\rm i}l_m)^{-2i}-(k_m^2-{\rm i}l_m)^{-2i}](k_m^4+l_m^2)^{2i}}{2k_m^{2i}}u_{n,\xi_{m0}}\\\label{ke}
&&\qquad\quad~=\sum_{m=1}^n (-1)^{i+1}\frac{{\rm i}[(k_m^2+{\rm i}l_m)^{-2i}-(k_m^2-{\rm i}l_m)^{-2i}](k_m^4+l_m^2)^{2i}}{2k_m^{2i}}\sigma_{\xi_{m0}}.
\end{eqnarray}
These two formulas universally characterize the action of all higher-order KP symmetries on $n$-soliton solutions, with the coefficients explicitly determined by the soliton parameters $k_m$ and $l_m$.
Remarkably, equations \eqref{ko} and \eqref{ke} for odd- and even-indexed symmetries can be cast into one compact unified closed-form expression valid for all integers $i\ge 1$
\begin{eqnarray*}
K_i\big|_{u=u_n}
&=&\sum_{m=1}^{n}
\frac{\mathrm{i}^{i+1}\big(k_m^2+\mathrm{i}l_m\big)^i
	+(-\mathrm{i})^{i+1}\big(k_m^2-\mathrm{i}l_m\big)^i}{2k_m^i}
\,u_{n,\xi_{m0}}\\
&=&\sum_{m=1}^{n}
\frac{\mathrm{i}^{i+1}\big(k_m^2+\mathrm{i}l_m\big)^i
	+(-\mathrm{i})^{i+1}\big(k_m^2-\mathrm{i}l_m\big)^i}{2k_m^i}
\,\sigma_{\xi_{m0}}.
\end{eqnarray*}

From relations \eqref{ko} and \eqref{ke}, it becomes clear that for the $n$-soliton solution $u = u_n$ given by \eqref{ns}, the infinite families of symmetries $K_{2i-1}$ and $K_{2i}$ are not independent. Indeed, only $n$ of them are linearly independent, and all remaining symmetries can be constructed as linear combinations of these $n$ fundamental ones. More explicitly, the restrictions of these symmetries to the $n$-soliton manifold admit compact determinant representations as follows:
	\begin{eqnarray*}\nonumber
		&&K_{2i-1}|_{u=u_n}=\sum_{m=1}^n\frac{{\rm i}}{2k_m^{2i-1}}\left[({\rm i}k_m^2+l_m)\left(-\frac{(k_m^4+l_m^2)^2}{(k_m^2+{\rm i}l_m)^2}\right)^{i-1}\right.\\\nonumber
		&&\qquad\qquad~~\left.\left.+({\rm i} k_m^2-l_m)\left(-\frac{(k_m^4+l_m^2)^2}{(k_m^2-{\rm i}l_m)^2}\right)^{i-1}\right]\frac{\Delta_{om}}{\Delta_o}\right|_{u=u_{n}},\\
		&&K_{2i}|_{u=u_n}=\left.\sum_{m=1}^n (-1)^{i+1}\frac{{\rm i}[(k_m^2+{\rm i}l_m)^{-2i}-(k_m^2-{\rm i}l_m)^{-2i}](k_m^4+l_m^2)^{2i}}{2k_m^{2i}}\frac{\Delta_{em}}{\Delta_e}\right|_{u=u_{n}},\\
		&&i=n+1, n+2,\dots.
	\end{eqnarray*}
	Here, $\Delta_o$ denotes the determinant of the $n \times n$ matrix $O$, i.e., $\Delta_o \equiv \det(O)$, with $O$ explicitly given by
	\begin{eqnarray*}\label{O}
	O= \begin{pmatrix}
	-k_1 & -k_2 & \cdots & -k_m & \cdots & -k_{n-1} & -k_n \vspace{2mm}\\
	\displaystyle\frac{k_1^4-3l_1^2}{k_1} & \displaystyle\frac{k_2^4-3l_2^2}{k_2} & \cdots & \displaystyle\frac{k_m^4-3l_m^2}{k_m} & \cdots & \displaystyle\frac{k_{n-1}^4-3l_{n-1}^2}{k_{n-1}} & \displaystyle\frac{k_n^4-3l_n^2}{k_n} \\
	\vdots & \vdots & \ddots & \vdots & \ddots & \vdots & \vdots \vspace{2mm}\\
	N_1 & N_2 & \cdots & N_m & \cdots & N_{n-1} & N_n
	\end{pmatrix},
	\end{eqnarray*}
	where the entries $N_i$ are defined as
	\begin{eqnarray*}
		N_i=\frac{{\rm i}}{2k_i^{2n-1}}\left[({\rm i}k_i^2+l_i)\left(-\frac{(k_i^4+l_i^2)^2}{(k_i^2+{\rm i}l_i)^2}\right)^{n-1}+({\rm i} k_i^2-l_i)\left(-\frac{(k_i^4+l_i^2)^2}{(k_i^2-{\rm i}l_i)^2}\right)^{n-1}\right],\quad i=1,\dots,n.
	\end{eqnarray*}

	Furthermore, $\Delta_{om}$ is defined as the determinant of the matrix $O_m$,  where $O_m$ is obtained from $O$ by replacing the $m$-th column with the vector $(K_1, K_3, \dots, K_{2n-1})^\mathrm{T}$, which collects the odd-order $K$-symmetries
	\begin{eqnarray*}\label{Om}
	O_m = \begin{pmatrix}
	-k_1 & -k_2 & \cdots & K_1 & \cdots & -k_{n-1} & -k_n \vspace{2mm}\\
	\displaystyle\frac{k_1^4-3l_1^2}{k_1} & \displaystyle\frac{k_2^4-3l_2^2}{k_2} & \cdots & K_3 & \cdots & \displaystyle\frac{k_{n-1}^4-3l_{n-1}^2}{k_{n-1}} & \displaystyle\frac{k_n^4-3l_n^2}{k_n} \vspace{2mm}\\
	\vdots & \vdots & \ddots & \vdots & \ddots & \vdots & \vdots \\
	N_1 & N_2 & \cdots & K_{2n-1} & \cdots & N_{n-1} & N_n
	\end{pmatrix}.
	\end{eqnarray*}
	The even-order counterparts are defined in complete analogy, with $\Delta_e \equiv \det(E)$ and $\Delta_{em} \equiv \det(E_m)$, where
	\begin{eqnarray*}\label{E}
		E = \begin{pmatrix}
			2l_1 & 2l_2 & \cdots & 2l_m & \cdots & 2l_{n-1} & 2l_n \vspace{2mm}\\
			\displaystyle\frac{4l_1(k_1^4-l_1^2)}{-k_1^2} & \displaystyle\frac{4l_2(k_2^4-l_2^2)}{-k_2^2} & \cdots & \displaystyle\frac{4l_m(k_m^4-l_m^2)}{-k_m^2} & \cdots & \displaystyle\frac{4l_{n-1}(k_{n-1}^4-l_{n-1}^2)}{-k_{n-1}^2} & \displaystyle\frac{4l_n(k_n^4-l_n^2)}{-k_n^2} \vspace{2mm}\\
			\vdots & \vdots & \ddots & \vdots & \ddots & \vdots & \vdots \\
			M_1 & M_2 & \cdots & M_m & \cdots & M_{n-1} & M_n
		\end{pmatrix},
	\end{eqnarray*}
	with the entries $M_i$ given by
	\begin{eqnarray*}
		M_i=(-1)^{n+1}\frac{{\rm i}[(k_i^2+{\rm i}l_i)^{-2n}-(k_i^2-{\rm i}l_i)^{-2n}](k_i^4+l_i^2)^{2n}}{2k_i^{2n}},\quad i=1,\dots,n.
	\end{eqnarray*}
	Finally, $E_m$ is obtained from $E$ by replacing the $m$-th column with the vector $(K_2, K_4, \dots, K_{2n})^\mathrm{T}$, which comprises the even-order $K$-symmetries
	\begin{eqnarray*}\label{Em}
		E_m = \begin{pmatrix}
			2l_1 & 2l_2 & \cdots & K_2 & \cdots & 2l_{n-1} & 2l_n \vspace{2mm}\\
			\displaystyle\frac{4l_1(k_1^4-l_1^2)}{-k_1^2} & \displaystyle\frac{4l_2(k_2^4-l_2^2)}{-k_2^2} & \cdots & K_4 & \cdots & \displaystyle\frac{4l_{n-1}(k_{n-1}^4-l_{n-1}^2)}{-k_{n-1}^2} & \displaystyle\frac{4l_n(k_n^4-l_n^2)}{-k_n^2}\vspace{2mm}\\
			\vdots & \vdots & \ddots & \vdots & \ddots & \vdots & \vdots \\
			M_1 & M_2 & \cdots & K_{2n} & \cdots & M_{n-1} & M_n
		\end{pmatrix}.
	\end{eqnarray*}

As an illustrative example, for the one-soliton solution $u_n = u_1$ given by \eqref{kp1}, there exist precisely two independent $K$-symmetries, which we may take to be $K_1$ and $K_2$. All higher-order symmetries are expressible as linear combinations of these two generators. More specifically,
	\begin{align*}
	K_{2i-1} &= \frac{\mathrm{i}}{-2k_1^{2i}}\Bigg[ (\mathrm{i}k_1^2 + l_1)\left(-\frac{(k_1^4 + l_1^2)^2}{(k_1^2 + \mathrm{i}l_1)^2}\right)^{\!i-1} \nonumber \\
	&\qquad\quad + (\mathrm{i}k_1^2 - l_1)\left(-\frac{(k_1^4 + l_1^2)^2}{(k_1^2 - \mathrm{i}l_1)^2}\right)^{\!i-1}\Bigg] K_1, \\[6pt]
	K_{2i} &= (-1)^{i+1}\frac{\mathrm{i}\big[(k_1^2 + \mathrm{i}l_1)^{-2i} - (k_1^2 - \mathrm{i}l_1)^{-2i}\big](k_1^4 + l_1^2)^{2i}}{4k_1^{2i}l_1}\,K_2.
	\end{align*}

For the two-soliton solution \eqref{kp2}, the corresponding relations become more intricate. For $i \ge 3$, the odd-order symmetries are given by
	\begin{align*}
	K_{2i-1} = \frac{\big[(k_2^4 - 3l_2^2)A_1 - (k_1^4 - 3l_1^2)A_2\big]K_1 + \big(k_2^2A_1 - k_1^2A_2\big)K_3}{k_1^4k_2^2 - k_1^2k_2^4 + 3k_1^2l_2^2 - 3k_2^2l_1^2},
	\end{align*}
where
	\begin{align*}
	A_j = \frac{\mathrm{i}}{2k_j^{2i-2}}\Bigg[ (\mathrm{i}k_j^2 + l_j)\left(-\frac{(k_j^4 + l_j^2)^2}{(k_j^2 + \mathrm{i}l_j)^2}\right)^{\!i-1} + (\mathrm{i}k_j^2 - l_j)\left(-\frac{(k_j^4 + l_j^2)^2}{(k_j^2 - \mathrm{i}l_j)^2}\right)^{\!i-1}\Bigg], \quad j = 1,2.
	\end{align*}
Analogously, the even-order symmetries take the form
	\begin{align*}
	K_{2i} = \frac{2\big[(k_2^4 - l_2^2)B_1 - (k_1^4 - l_1^2)B_2\big]K_2 + \big(k_2^2B_1 - k_1^2B_2\big)K_4}{2\big(k_1^4k_2^2 - k_1^2k_2^4 + k_1^2l_2^2 - k_2^2l_1^2\big)}, \qquad i \ge 3,
	\end{align*}
with the coefficients defined by
	\begin{align*}
	B_j = (-1)^i\,\frac{\mathrm{i}\big[(k_j^2 + \mathrm{i}l_j)^{-2i} - (k_j^2 - \mathrm{i}l_j)^{-2i}\big](k_j^4 + l_j^2)^{2i}}{4k_j^{2i-2}l_j}, \quad j = 1,2.
	\end{align*}

The expressions explicitly characterize the reduction of the $K$-symmetry hierarchy to the $n$-soliton manifold. They reveal that, despite the infinite-dimensional nature of the full symmetry algebra, its restriction to a given $n$-soliton solution is completely governed by the finite set of parameters $\{k_m, l_m, \xi_{m0}\}$ and admits a concise algebraic description in terms of determinants and dispersion relations. This observation naturally leads to the question of whether analogous  formulas exist for the $\tau$-symmetries. 

For the $\tau$-symmetries, explicit restrictions to the $n$-soliton solution \eqref{ns} can also be obtained. Up to the eighth order, they read
	\begin{eqnarray*}
		&&\tau_2|_{u=u_n}=\sum_{m=1}^n(-k_m\sigma_{l_m}),\\
		&&\tau_3|_{u=u_n}=\sum_{m=1}^n\left(k_m\sigma_{k_m}+2l_m\sigma_{l_m}\right),\\
		&&\tau_4|_{u=u_n}=\sum_{m=1}^n\left(-2l_m\sigma_{k_m}+\frac{k_m^4-3l_m^2}{k_m}\sigma_{l_m}\right),\\
		&&\tau_5|_{u=u_n}=\sum_{m=1}^n\left(-\frac{k_m^4-3l_m^2}{k_m}\sigma_{k_m}-\frac{4l_m(k_m^4-l_m^2)}{k_m^2}\sigma_{l_m}+2k_m^2\sigma_{\xi_{m0}}\right),\\
		&&\tau_6|_{u=u_n}=\sum_{m=1}^n\left(\frac{4l_m(k_m^4-l_m^2)}{k_m^2}\sigma_{k_m}-\frac{k_m^8-10k_m^4l_m^2+5l_m^4}{k_m^3}\sigma_{l_m}-8k_ml_m\sigma_{\xi_{m0}}\right),\\
		&&\tau_7|_{u=u_n}=\sum_{m=1}^n\left(\frac{k_m^8-10k_m^4l_m^2+5l_m^4}{k_m^3}\sigma_{k_m}+\frac{2l_m(3k_m^8-10k_m^4l_m^2+3l_m^4)}{k_m^4}\sigma_{l_m}-4(k_m^4-5l_m^2)\sigma_{\xi_{m0}}\right),\\
		&&\tau_8|_{u=u_n}=\sum_{m=1}^n\left(-\frac{2l_m(3k_m^8-10k_m^4l_m^2+3l_m^4)}{k_m^4}\sigma_{k_m}+\frac{k_m^{12}-21k_m^8l_m^2+35k_m^4l_m^4-7l_m^6}{k_m^5}\sigma_{l_m}\right.\\&&\qquad\qquad~\left.+\frac{8l_m(3k_m^4-5l_m^2)}{k_m}\sigma_{\xi_{m0}}\right).
	\end{eqnarray*}
Each $\tau_n$ is a linear combination of the three fundamental soliton symmetries $\sigma_{k_m}$, $\sigma_{l_m}$ and $\sigma_{\xi_{m0}}$, with coefficients that are rational functions of $k_m$ and $l_m$. 
The coefficients in the series admit a unified recursive construction. Define the decomposition
\begin{eqnarray*}
\tau_{j}|_{u=u_n}=\sum_{m=1}^{n}\big(A_{j,m}\sigma_{k_{m}}+B_{j,m}\sigma_{l_{m}}+C_{j,m}\sigma_{\xi_{m0}}\big),\quad j\ge 2.
\end{eqnarray*}
The coefficients satisfy the relation \(A_{j+1,m}=-B_{j,m}\). Both \(A_{j,m}\) and \(B_{j,m}\) obey the linear recurrence
\begin{eqnarray*}
A_{j+2,m}=-\frac{2l_{m}}{k_{m}}A_{j+1,m}-\left(k_{m}^{2}+\frac{l_{m}^{2}}{k_{m}^{2}}\right)A_{j,m},
\end{eqnarray*}
with initial conditions \(A_{2,m}=0\), \(A_{3,m}=k_{m}\).
Equivalently, \(A_{j,m}\) admits the closed form expression
\begin{equation*}
A_{j,m} = -\frac{\left({\rm i}(l_m^2- k_m^4)  - 2 k_m^2 l_m\right)\left(\dfrac{-k_m^4-l_m^2}{\left({\rm i} k_m^2+l_m\right)\,k_m}\right)^{\!j}
	+\left({\rm i} (k_m^4 -  l_m^2) - 2 k_m^2 l_m\right)\left(\dfrac{k_m^4+l_m^2}{\left({\rm i} k_m^2-l_m\right)\,k_m}\right)^{\!j}}{2\left(k_m^4+l_m^2\right)^2}\,k_m^2.
\end{equation*}
The phase-offset coefficients are determined by
\begin{eqnarray*}
C_{j,m}=-\frac{j-3}{k_{m}}A_{j,m}-\frac{(j-2)l_{m}}{k_{m}^{2}}A_{j-1,m},\quad j\ge 3,\qquad C_{2,m}=0.
\end{eqnarray*}

Introduce the vector of elementary tangent symmetries
\begin{equation*}
\boldsymbol{\sigma}
=
\left(
\sigma_{k_1},\sigma_{l_1},\sigma_{\xi_{10}},
\ldots,
\sigma_{k_n},\sigma_{l_n},\sigma_{\xi_{n0}}
\right)^{\mathsf T},
\end{equation*}
and define the coefficient row vector
\begin{equation*}
\mathbf{q}_i
=
\left(
A_{i,1},B_{i,1},C_{i,1},
\ldots,
A_{i,n},B_{i,n},C_{i,n}
\right).
\end{equation*}
We select the first $3n$ restricted $\tau$-symmetries, $\tau_3, \tau_4, \ldots, \tau_{3n+2}$,
and form the $3n \times 3n$ coefficient matrix
\begin{equation*}
M_n
=
\begin{pmatrix}
\mathbf{q}_3\\
\mathbf{q}_4\\
\vdots\\
\mathbf{q}_{3n+2}
\end{pmatrix}
=
\begin{pmatrix}
A_{3,1} & B_{3,1} & C_{3,1} & \cdots & A_{3,n} & B_{3,n} & C_{3,n}\\
A_{4,1} & B_{4,1} & C_{4,1} & \cdots & A_{4,n} & B_{4,n} & C_{4,n}\\
\vdots  & \vdots  & \vdots  & \ddots & \vdots  & \vdots  & \vdots\\
A_{3n+2,1} & B_{3n+2,1} & C_{3n+2,1} & \cdots & A_{3n+2,n} & B_{3n+2,n} & C_{3n+2,n}
\end{pmatrix}.
\end{equation*}
Let $\Delta_n = \det(M_n)$, and assume $\Delta_n \neq 0$.
For a given $i \ge 3n+3$, let $M_{n,i}^{(r)}$ denote the matrix obtained by replacing the
$r$-th row of $M_n$ by the coefficient row vector $\mathbf{q}_i$.
Then, by Cramer's rule, the higher-order restricted $\tau$-symmetry can be written as
\begin{equation*}
\tau_i
=
\sum_{r=1}^{3n}
\frac{\det\!\left(M_{n,i}^{(r)}\right)}{\Delta_n}
\tau_{r+2}.
\end{equation*}
This formula shows that, on the finite-dimensional $n$-soliton manifold, every higher-order
$\tau$-symmetry is a linear combination of the first $3n$ restricted $\tau$-symmetries.
The determinant representation is a direct consequence of the finite-dimensionality of the
tangent space and Cramer's rule.

For $n=1$, the coefficient matrix in the general formula becomes
\begin{equation*}
M_1
=
\begin{pmatrix}
A_{3,1} & B_{3,1} & C_{3,1}\\
A_{4,1} & B_{4,1} & C_{4,1}\\
A_{5,1} & B_{5,1} & C_{5,1}
\end{pmatrix}
=
\begin{pmatrix}
k_1 & 2l_1 & 0\\
-2l_1 & q & 0\\
-q & r & 2k_1^2
\end{pmatrix},
\end{equation*}
where the dispersion-related quantities are
\begin{equation*}
q = \frac{k_1^4-3l_1^2}{k_1},
\qquad
r = -\frac{4l_1(k_1^4-l_1^2)}{k_1^2}.
\end{equation*}
The restricted $\tau$-symmetries satisfy
\begin{equation*}
\begin{pmatrix}
\tau_3\\
\tau_4\\
\tau_5
\end{pmatrix}
= M_1
\begin{pmatrix}
\sigma_{k_1}\\
\sigma_{l_1}\\
\sigma_{\xi_{10}}
\end{pmatrix}.
\end{equation*}
The next restricted symmetry $\tau_6$ has the coefficient row
$\mathbf{q}_6 = (A_{6,1}, B_{6,1}, C_{6,1})$.
If $\Delta_1 = \det(M_1) \neq 0$, then by Cramer's rule we have the explicit representation
\begin{equation*}
\tau_6
=
\frac{\det\!\begin{pmatrix} A_{6,1} & B_{6,1} & C_{6,1} \\ -2l_1 & q & 0\\
	-q & r & 2k_1^2  \end{pmatrix}}{\Delta_1}
\tau_3
+
\frac{\det\!\begin{pmatrix} k_1 & 2l_1 & 0 \\ A_{6,1} & B_{6,1} & C_{6,1} \\ -q & r & 2k_1^2 \end{pmatrix}}{\Delta_1}
\tau_4
+
\frac{\det\!\begin{pmatrix} k_1 & 2l_1 & 0\\
	-2l_1 & q & 0 \\ A_{6,1} & B_{6,1} & C_{6,1} \end{pmatrix}}{\Delta_1}
\tau_5.
\end{equation*}
Thus, whenever $\Delta_1 \neq 0$, the fourth restricted symmetry $\tau_6$
is uniquely expressed as a linear combination of $\tau_3$,
$\tau_4$, and $\tau_5$, illustrating the general
$3n$-dimensional representation at $n=1$.

\subsection{Tangent space of the multi-soliton manifold}

From our explicit reduction formulas, one observes a clear geometric dichotomy between $K$-symmetries and $\tau$-symmetries. When restricted to $M_n$, each commuting $K$-symmetry generator $K_i$ lies entirely within the $n$-dimensional subspace spanned solely by phase‑shift generators $\{\sigma_{\xi_{m0}}\}$. In other words, isospectral $K$-symmetry flows only induce phase translations for individual solitons without modifying wave‑number parameters $k_m$ or direction parameters $l_m$. In contrast, the master‑symmetry originated $\tau$-symmetries, which correspond to non‑isospectral transformations, possess non‑vanishing projections onto all three types of tangent‑space basis vectors $\sigma_{k_m},\sigma_{l_m},\sigma_{\xi_{m0}}$. They are capable of deforming soliton wave numbers, direction angles as well as soliton phases.

There exists a natural restriction homomorphism from the infinite‑dimensional symmetry Lie algebra $\mathfrak{g}_\infty$ onto the finite‑dimensional tangent space $T_{u_n}M_n$:
\[
\rho:\mathfrak{g}_\infty \to T_{u_n}M_n.
\]
The kernel $\ker(\rho)$ is non‑trivial: infinitely many higher‑order symmetry generators are mapped to zero modulo the tangent directions of $M_n$. This degeneracy explains why higher‑order symmetry flows for $i>n$ can be expressed as linear combinations of the first $n$ independent restricted generators, as represented by the determinant formulas derived in the previous subsection.

\section{Extension to the mKP equation: symmetry reduction on multi-soliton manifolds}\label{mkph}

The mKP equation emerges from the KP equation via a Miura-type transformation. Introducing the potential field $v$ through
\begin{equation}\label{mt}
u = \mathrm{i}v_x + v^2 + \partial_x^{-1}v_y, 
\end{equation}
a direct substitution verifies that if $u$ satisfies the KP equation \eqref{kp}, then $v$ defined by \eqref{mt} indeed satisfies the mKP equation
\begin{equation*}\label{mkp}
v_t = v_{xxx} + 6v^2 v_x + 6v_x \partial_x^{-1} v_y - 3 \partial_x^{-1} v_{yy} . 
\end{equation*} Conversely, the mKP equation can be interpreted as a modification of the KP equation, analogous to the relationship between the KdV and mKdV equations in (1+1) dimensions.


Following the theoretical framework established for the KP equation, infinite commuting symmetry flows can be explicitly constructed for the mKP equation. Representative generators of these symmetry flows are presented below
	\[
	\begin{aligned}
	K_1 &= -v_x ,\\
	K_2 &= 2 v_y ,\\
	K_3 &= v_{xxx} + 6v^2 v_x + 6v_x \partial_x^{-1} v_y - 3 \partial_x^{-1} v_{yy} ,\\
	K_4 &= -4 \big( v_{xxy} + 4v^2 v_y - 2 \partial_x^{-1} v_y^2 - \partial_x^{-2} v_{yyy} + 4 v_y \partial_x^{-1} v_y + 4 v_x \partial_x^{-1} (v v_y) \\
	&\quad + 2 v_x \partial_x^{-2} v_{yy} + 2 v \partial_x^{-1} v_{yy} - 2 \partial_x^{-1} (v v_{yy}) \big) , \\
	K_5 &= -v_{5x} + 10 v_{xyy} - 10 v_{xx} v_y - 6 (v^5)_x - 10 v^2 v_{xxx} - 10 v_x^3 - 10 v_x v_{xy} \\
	&\quad - 20 v (v_x^2)_x - 5 \partial_x^{-3} v_{4y} + 10 v \partial_x^{-2} v_{yyy} + 10 v^2 \partial_x^{-1} v_{yy} \\
	&\quad + 20 v \big( \partial_x^{-1} v_y^2 + \partial_x^{-1} (v v_{yy}) \big) - 15 \partial_x^{-2} (v_y^2)_y - 10 \partial_x^{-2} (v v_{yyy}) \\
	&\quad + 10 \big( \partial_x^{-3} v_{yyy} + 2 \partial_x^{-2} (v v_{yy}) + 2 \partial_x^{-2} (v_y^2) - 3 (\partial_x^{-1} v_y)^2 \big) v_x \\
	&\quad + 20 \big( \partial_x^{-2} v_{yy} + \partial_x^{-1} (v^2)_y \big) v_y - 10 \big( v_{xxx} + 2 (v^3)_x - 3 \partial_x^{-1} v_{yy} \big) \partial_x^{-1} v_y .
	\end{aligned}
	\]
	These symmetry generators satisfy the commutation relations $[K_m, K_n] = 0$ for all positive integers $m, n$, thereby confirming the integrability of the mKP hierarchy.
	The higher-order $K_i$ symmetries can also be recursively computed via
	\begin{eqnarray*}
	K_{n+1}=\frac{1}{n} [\tau_3, K_n], \quad n \geq 1,
	\end{eqnarray*} 
	where the base $\tau_3$ symmetry is given by
	\begin{eqnarray*}
	\tau_3=3tK_4+yK_3-xK_2-v^2-3\partial_{x}^{-1}v_y.
	\end{eqnarray*}
	The full family of $\tau$-symmetries is further generated through the commutator recursion \cite{ybib23}
	\begin{eqnarray}\label{mkpt}
	\tau_{n+1}=\frac{1}{n-3} [\tau_3, \tau_n], \quad n \neq 3.
	\end{eqnarray} 
	Several low-order $\tau$-symmetries  read
	\begin{align*}
	&\tau_1=3tK_2+yK_1+\frac12,\\
	&\tau_2=3tK_3+yK_2-xK_1+v,\\
	&\tau_3=3tK_4+yK_3-xK_2-v^2-3\partial_{x}^{-1}v_y,\\
	&\tau_4=3tK_5+yK_4-xK_3-4v^3-3v_{xx}-2v_x(\partial_{x}^{-1}v^2+\partial_{x}^{-2}v_y)-8v\partial_{x}^{-1}v_y\\&\quad+5\partial_{x}^{-2}v_{yy}+10\partial_{x}^{-1}(vv_y).
	\end{align*}

When restricted to the $n$-soliton background of the mKP equation, given by
\begin{equation}\label{mkpns}
v_n=-\mathrm{i}(\ln f)_x,~ f=1+\sum_{i=1}^n\exp(\xi_i),~ \xi_i=k_ix+\mathrm{i}k_i^2y+4k_i^3t+\xi_{i0},
\end{equation}
every symmetry generator collapses to a linear combination of the elementary phase‑shift symmetries \(\sigma_{\xi_{m0}} = v_{n,\xi_{m0}}\) for \(m=1,2,\dots,n\). The corresponding explicit reduction relations are 
	\begin{eqnarray*}
		&&K_1|_{v=v_n}=-\sum_{m=1}^nk_mv_{n,\xi_{m0}},\\
		&&K_2|_{v=v_n}=2{\rm i}\sum_{m=1}^nk_m^2v_{n,\xi_{m0}},\\
		&&K_3|_{v=v_n}=4\sum_{m=1}^nk_m^3v_{n,\xi_{m0}},\\
		&&K_4|_{v=v_n}=-8{\rm i}\sum_{m=1}^nk_m^4v_{n,\xi_{m0}},\\
		&&K_5|_{v=v_n}=-16\sum_{m=1}^nk_m^5v_{n,\xi_{m0}},\\
		&&K_6|_{v=v_n}=32{\rm i}\sum_{m=1}^nk_m^6v_{n,\xi_{m0}},\\
		&&\vdots
	\end{eqnarray*}
	More compactly, for all $i = 1,2,\ldots,\infty$, we obtain
	\begin{align*}
	K_{i}|_{v=v_n}=-(-2{\rm i})^{i-1} \sum_{m=1}^nk_m^iv_{n,\xi_{m0}}.
	\end{align*}	
	Equivalently, although the mKP equation possesses infinitely many $K$-symmetries, only $n$ generators are linearly independent when restricted onto the $n$-soliton manifold. Remaining higher‑order symmetries can be expressed as their linear combinations
	\begin{eqnarray*}
	K_i|_{v=v_n}=-(-2{\rm i})^{i-1} \sum_{m=1}^nk_m^i\frac{\Delta_{nm}}{\Delta_n} ,~ i=n+1, n+2, \dots, \infty,
	\end{eqnarray*}
	where $\Delta_n$ denotes the determinant of the $n\times n$ matrix $N$
	\begin{eqnarray*}\label{N}
		N = \begin{pmatrix}
			-k_1 & -k_2 & \cdots & -k_m & \cdots & -k_{n-1} & -k_n \vspace{2mm}\\
			\displaystyle2{\rm i}k_1^2 & \displaystyle2{\rm i}k_2^2 & \cdots & 2{\rm i}k_m^2 & \cdots & \displaystyle2{\rm i}k_{n-1}^2 & \displaystyle2{\rm i}k_n^2\vspace{2mm}\\
			\vdots & \vdots & \ddots & \vdots & \ddots & \vdots & \vdots \\
			-(-2{\rm i})^{n-1}k_1^n & -(-2{\rm i})^{n-1}k_2^n & \cdots & -(-2{\rm i})^{n-1}k_m^n & \cdots & -(-2{\rm i})^{n-1}k_{n-1}^n & -(-2{\rm i})^{n-1}k_n^n
		\end{pmatrix},
	\end{eqnarray*}
	and $\Delta_{nm}$ stands for the determinant of matrix $N_m$
	\begin{eqnarray*}\label{Nm}
		N_m = \begin{pmatrix}
			-k_1 & -k_2 & \cdots & K_1 & \cdots & -k_{n-1} & -k_n \vspace{2mm}\\
			\displaystyle2{\rm i}k_1^2 & \displaystyle2{\rm i}k_2^2 & \cdots & K_2 & \cdots & \displaystyle2{\rm i}k_{n-1}^2 & \displaystyle2{\rm i}k_n^2\vspace{2mm}\\
			\vdots & \vdots & \ddots & \vdots & \ddots & \vdots & \vdots \\
			-(-2{\rm i})^{n-1}k_1^n & -(-2{\rm i})^{n-1}k_2^n & \cdots & K_n & \cdots & -(-2{\rm i})^{n-1}k_{n-1}^n & -(-2{\rm i})^{n-1}k_n^n
		\end{pmatrix}.
	\end{eqnarray*}

	Building upon the preceding discussion of $K$-symmetries, we next investigate the restriction behaviors of \(\tau\)-symmetries \eqref{mkpt} on $n$-soliton background \eqref{mkpns}. 
Several low‑order restricted $\tau$-symmetries are
\begin{align*}
&\tau_2|_{v=v_n}=\sum_{m=1}^nk_mv_{n,k_m},\\
&\tau_3|_{v=v_n}=-2{\rm i}\sum_{m=1}^nk_m^2v_{n,k_m}-{\rm i}\sum_{m=1}^nk_mv_{n,\xi_{m0}},\\
&\tau_4|_{v=v_n}=-4\sum_{m=1}^nk_m^3v_{n,k_m}-2\sum_{m=1}^nk_m^2v_{n,\xi_{m0}},\\
&\tau_5|_{v=v_n}=8{\rm i}\sum_{m=1}^nk_m^4v_{n,k_m}+4{\rm i}\sum_{m=1}^nk_m^3v_{n,\xi_{m0}}.
\end{align*}
For arbitrary order $i>2$, we obtain the closed‑form formula
\begin{align*}
&\tau_{i}=(-2{\rm i})^{i-2}\sum_{m=1}^nk_m^{i-1}v_{n,k_m}+\frac{(-2{\rm i})^{i-2}}{2}\sum_{m=1}^nk_m^{i-2}v_{n,\xi_{m0}}.
\end{align*}
When $i>n+1$, higher‑order restricted $\tau$-symmetries can be expressed via determinant expansions
\begin{align*}
&\tau_{i}=2(-1)^n(-2{\rm i})^{i-4}h_{i-n-2}\prod_{j=1}^nk_j\sum_{m=1}^n\frac{\Delta_{nm}}{\Delta_n}-(-2{\rm i})^{i-2}\sum_{m=1}^nk_m^{i-1}\frac{\Delta_{tm}}{\Delta_n}.
\end{align*}
Here $h_{i}$ denote complete homogeneous symmetric polynomials, whose generating function is
\begin{equation*}
\sum_{i=0}^{\infty} h_i t^i=\prod_{j=1}^{n}\frac{1}{1-k_j t}.
\end{equation*}
In the above expression, $\Delta_{tm}$ is the determinant of matrix $T_{m}$ defined by
	\begin{eqnarray*}\label{Tm}
	T_m = \begin{pmatrix}
		-k_1 & -k_2 & \cdots & \tau_2 & \cdots & -k_{n-1} & -k_n \vspace{2mm}\\
		\displaystyle2{\rm i}k_1^2 & \displaystyle2{\rm i}k_2^2 & \cdots & \tau_3 & \cdots & \displaystyle2{\rm i}k_{n-1}^2 & \displaystyle2{\rm i}k_n^2\vspace{2mm}\\
		\vdots & \vdots & \ddots & \vdots & \ddots & \vdots & \vdots \\
		-(-2{\rm i})^{n-1}k_1^n & -(-2{\rm i})^{n-1}k_2^n & \cdots & \tau_{n+1} & \cdots & -(-2{\rm i})^{n-1}k_{n-1}^n & -(-2{\rm i})^{n-1}k_n^n
	\end{pmatrix}.
\end{eqnarray*}
The determinants $\Delta_{n}$ and $\Delta_{nm}$ employ the identical matrix construction as that adopted for the mKP $K$-symmetry reductions in the preceding paragraph.

To further validate the self‑consistency of the above symmetry reduction results, we perform the dimensional reduction \(v_y=0\). Under this constraint, the (2+1)-dimensional mKP equation degenerates into the well‑known (1+1)-dimensional mKdV equation \cite{dbib8}
\begin{equation*}
v_{t}=v_{xxx}+6v^{2}v_{x}.
\end{equation*}
Correspondingly, the hierarchy of symmetry generators \(K_i\) contracts to the standard symmetry flows of the mKdV equation. Specifically, odd‑order flows \(K_{2i-1}\) recover the complete mKdV hierarchy \cite{25pla}, while all even‑order flows \(K_{2i}\) vanish identically after this reduction. Such consistent reduction guarantees that our (2+1)-dimensional symmetry framework faithfully reproduces established results for the integrable mKdV equation in lower dimensions.

\section{Conclusions and discussions}\label{con}

In this work, we have investigated how infinite‑dimensional symmetry algebras manifest themselves when restricted onto finite‑dimensional multi‑soliton manifolds for two representative (2+1)-dimensional integrable hierarchies, the KP and mKP equations. Explicit computations verify a dispersion‑driven reduction mechanism: each commuting $K$-symmetry generator, after restriction to an $n$-soliton background, becomes a linear combination of elementary phase shift generators, with expansion coefficients exactly given by dispersion relations of the corresponding hierarchy.

For the KP equation, closed‑form reduction formulas for arbitrary order odd‑ and even‑indexed $K$-symmetries are derived, together with determinant representations reconstructing higher‑order restricted symmetries from a finite set of lower‑order basis generators. For $\tau$-symmetries, each restricted generator decomposes into a linear combination of three fundamental soliton deformations: variations of wave numbers, direction parameters and phase offsets. Analogous reduction patterns are established for the mKP hierarchy, whose simpler monomial spectral coefficients stem from its particular dispersion relation.

Our results provide a unified physical picture for understanding abstract infinite symmetries acting on multi‑soliton states. Higher‑order infinite symmetries do not produce entirely new soliton configurations. Instead, their net effect on multi‑soliton solutions is to generate weighted superpositions of individual soliton phase translations. Each constituent soliton responds to symmetry transformations according to its own dispersion law, and multi‑soliton interaction effects do not destroy this decomposition property.
From the wave physics perspective, higher‑order commuting symmetry flows correspond to higher‑order dispersive evolutionary flows. When acting on multi‑soliton backgrounds, these higher‑order dispersive flows effectively induce weighted phase drifts for each individual soliton component. This interpretation bridges abstract algebraic symmetry structures and observable dynamical behaviors of coherent nonlinear wave configurations.

Several limitations of the present work should be stressed. Our analysis is confined to standard Hirota‑type line $n$-soliton solutions, and it remains unclear whether the same dispersion‑driven reduction mechanism persists for lump solutions, rational solutions, or mixed soliton‑lump configurations. Looking ahead, there are several promising avenues worthy of further exploration. It would be valuable to examine whether symmetry reduction outcomes can be directly deduced solely from Lax‑pair information together with dispersion relations, circumventing the heavy commutator‑based recursion computations required to generate high‑order symmetry generators. Another interesting direction is to investigate whether the recurrence relations satisfied by restricted $\tau$-symmetries on multi‑soliton manifolds can be utilized to build action‑angle‑type variables, thereby generalizing the (1+1)-dimensional geometric framework established by Fuchssteiner and Oevel \cite{bbib16}. Furthermore, a rigorous investigation into how symmetry‑reduction structures transform under the Miura‑type map connecting the mKP and KP equations would deepen our understanding of the algebraic interplay between these two hierarchies. Further investigations along these lines will help clarify to what extent dispersion‑driven reduction can serve as a general feature for soliton‑bearing integrable partial differential equations.

\section*{Acknowledgement}
The work was sponsored by the National Natural Science Foundation of China (Nos. 12301315,  12235007), the Natural Science Foundation of Zhejiang Province No. LMS25A010002.
\\\\	
\noindent \bf Data Availability
\rm No datasets were generated or analyzed during the current study.
\section*{\large Declarations}
\noindent \bf Conflict of interest \rm The authors declare no conflict of interest.

\bibliographystyle{elsarticle-num}
\bibliography{ref}

\end{document}